\documentclass[final,5p,times,twocolumn]{elsarticle}

\usepackage{amssymb}
\usepackage{lipsum}

\usepackage{amsmath}
\usepackage{color}

\journal{Physics Letters B}

\usepackage{graphicx}
\usepackage{dcolumn}
\usepackage{bm}
\usepackage{braket}
\usepackage{amsmath}
\usepackage{here}
\usepackage{color}
\usepackage{comment}

\usepackage[small]{caption}
\usepackage[subrefformat=parens]{subcaption}
\newcommand{\Slash}[1]{{\ooalign{\hfil/\hfil\crcr$#1$}}}

\def\slashb#1{\not\!\!#1}

\begin{document}

\begin{frontmatter}

\title{
Spontaneous chiral-symmetry breaking 
in four-dimensional QCD \\
with $A_x=A_y=0$ projection 
in dimensional reduction gauge
}

\author[label1]{Kei Tohme}
\affiliation[label1]{organization={Department of Physics, Kyoto University},
            addressline={Kitashirakawaoiwake, Sakyo}, 
            city={Kyoto},
            postcode={606-8502}, 
            state={Kyoto},
            country={Japan}}
            
\author[label1]{Hideo Suganuma}

\begin{abstract}
We investigate spontaneous chiral-symmetry breaking in 
four-dimensional (4D) QCD 
in the dimensional reduction (DR) gauge, defined so as to minimize
$R_{\mathrm{DR}}\equiv~\int d^{4}s ~ \mathrm{Tr} \left[ A_{x}^{2}(s) + A_{y}^{2}(s) \right]$.
We consider ``$tz$-projection" of $A_{x,y}(s) \to 0$ 
and ``$xy$-projection" of $A_{t,z}(s) \to 0$, respectively, 
for the gauge configuration generated in the DR gauge, in a similar sense to Abelian, monopole and photon  projections in the maximally Abelian gauge. 
The chiral condensate is estimated from the 
low-lying Dirac-eigenvalue density $\rho(\lambda)$ 
using the Banks-Casher relation. 
By the $tz$-projection in the DR gauge, 
the chiral condensate is largely reserved. 
In contrast, 
the chiral condensate becomes zero  
by the $xy$-projection in the DR gauge.
This indicates that $A_{t}(s)$ and $A_{z}(s)$ play a dominant role, while $A_{x}(s)$ and $A_{y}(s)$ play 
no role, in spontaneous chiral-symmetry breaking
in the DR gauge.
\end{abstract}


\end{frontmatter}

\section{Introduction}

Since 1974, quantum chromodynamics (QCD) has been  
established as the fundamental theory of strong interactions.
While high-energy hadron reaction can be 
treated by pertuebative QCD, 
it is highly difficult to solve low-energy QCD analytically due to its strong-coupling nature. 
Instead, since 1980, lattice QCD Monte Carlo calculations have been applied as a non-perturbative analysis \cite{C80,Rothe}. 

In the real world, 
four-dimensional (4D) QCD exhibits 
two important non-perturbative 
properties of color confinement \cite{Greensite} 
and spontaneous chiral-symmetry breaking \cite{NJL61}.
The chiral symmetry relates to quasi-massless 
current quarks in QCD in even-dimensional space-time.
In 4D QCD, the chiral symmetry is spontaneously broken, and its order parameter is 
the chiral condensate $\langle \bar qq \rangle$.

In 4D QCD, quark confinement is characterized by a linear interquark potential  \cite{Handbook}
and one-dimensional squeezing of color-electric fields, which is idealized as the string picture in the infrared region. 
Here, the string tension $\sigma \simeq 0.89 {\rm GeV/fm}$ gives the quark confining force and is a key parameter of quark confinement.
The formation of one-dimensional color-flux-tube is observed between (anti)quarks in lattice QCD calculations \cite{Rothe, IBSS03, Bicudo, Bicudo2}. 

As an interesting conjecture, 
the one-dimensional flux-tube formation can be regarded as an effective dimensional reduction of spatial dimensions by two, 
similar to the Parisi-Sourlas mechanism \cite{Parisi} for the equivalence
of a $d$-dimensional spin system under Gaussian random magnetic fields and 
a $(d-2)$-dimensional system without magnetic fields.
In fact, the nonperturbative QCD vacuum 
might have a dimensional reduction nature \cite{Kondo, Iritani}, or a dimensional reduction approach might be useful \cite{DR_Greensite}. 

In our previous paper \cite{TS24}, 
we have considered the possibility of describing the 4D Yang-Mills theory in terms of 
the two-dimensional (2D) gluon degrees of freedom, 
using the dimensional reduction (DR) gauge,  
which is defined so as to minimize
$R_{\mathrm{DR}}\equiv~\int d^{4}s ~ \mathrm{Tr} \left[ A_{x}^{2}(s) + A_{y}^{2}(s) \right]$.
From the lattice QCD analysis in the DR gauge, 
the gluon propagation in the $x$ and $y$
directions is suppressed, 
and gluons are bounded on $t$-$z$ planes 
in DR-gauged QCD. 
Since the amplitude of $A_x$ and $A_y$ is strongly suppressed in the DR gauge,  
the $tz$-projection of $A_{x,y}(s) \to 0$ 
is performed for the gauge configuration generated in lattice QCD in the DR gauge. 
By the $tz$-projection, 
the inter-quark potential remains unchanged, 
which suggests that 
$A_{t}$ and $A_{z}$ play a dominant role in quark confinement in the DR gauge \cite{TS24}.

However, one might suspect that 
the dimensional reduction is compatible with 
spontaneous chiral-symmetry breaking. 
For, the Coleman theorem \cite{Coleman, Mermin-Wagner} prohibits   
spontaneous breaking of global continuous symmetry 
in 1+1 dimensional systems, because 
the appearance of massless Nambu-Goldstone bosons causes an infrared instability.

In this Letter, similar to the 
previous study on quark confinement \cite{TS24,TS25}, we investigate 
spontaneous chiral-symmetry breaking,  
which is another important non-perturbative quantity of 4D QCD, 
in the context of effective dimensional reduction, 
using the DR gauge and the $tz$-projection.
Here, we consider that 
spontaneous chiral-symmetry breaking is compatible with 
the gluonic dimensional reduction 
in DR-gauged QCD.

This Letter is organized as follows.
In Sec.\ref{sec:DR gauge}, we introduce the DR gauge fixing and $tz$-projection in both continuum and lattice QCD.
In Sec.\ref{sec:chiral}, 
we briefly review spontaneous chiral-symmetry breaking and the Banks-Casher relation \cite{BC80}.
In Sec.\ref{sec:numerical}, we perform the lattice QCD calculations and numerical analyses of spontaneous chiral-symmetry breaking in the DR gauge.
Section \ref{sec:discussion} is 
devoted for the discussion.  
Section \ref{sec:summary} is devoted for the summary and concluding remarks.

\section{Dimensional reduction gauge in QCD}
\label{sec:DR gauge}

In this section, we introduce the DR gauge 
together with the $tz$- and $xy$-projections 
in 4D SU($N_c$) QCD \cite{TS24,TS25}.
In 4D Minkowski space-time, 
the SU($N_c$) QCD action is expressed 
by the gluon field $A_\mu(s) \in {\rm su}(N_c)$ and the quark field $q$:
\begin{eqnarray}
    \label{eq:QCD_action}
    S_{\mathrm{QCD}}= 
    \int d^4 s~
    \Bigg[-
    \frac{1}{2}
    \mathrm{Tr} \; G_{\mu\nu}G^{\mu\nu}
    +
    \bar q(i\Slash{D}-m)q \Bigg].
\end{eqnarray}
We have used 
the covariant derivative 
$D_\mu\equiv \partial_\mu+igA_\mu$ 
and the field strength
$G_{\mu\nu}\equiv \frac{1}{ig}[D_\mu, D_\nu]
=\partial_\mu A_\nu-\partial_\nu A_\mu+ig[A_\mu, A_\nu]$, 
with the QCD gauge coupling $g$.
Here, $m$ denotes the current quark mass, 
and its small value for u, d quarks 
leads to global chiral symmetry.
In this Letter, $s = (x,y,z,t)$ denotes 
a four-vector of space-time coordinates.

\subsection{Dimensional reduction gauge}

For the gluon field 
$A_\mu(s)$ in QCD, 
the DR gauge is defined so as to  
minimize 
\begin{eqnarray}
    \label{eq:DRG}
    R_{\mathrm{DR}}[A_x,A_y]
     \equiv 
    \int \hspace{-2pt}
    d^{4} s \;
    \mathrm{Tr}\left[ 
    A_{x}(s)^2 + A_{y}(s)^2
    \right] 
\end{eqnarray}
by the SU($N_c$) gauge transformation.
This global definition of the DR gauge 
can be applied to both Minkowski and Euclidean space-times, 
since the DR gauge fixing can be performed locally in the temporal direction 
like the Coulomb gauge.

Note that 
DR-gauged QCD has a residual gauge symmetry 
with the gauge function $\Omega(t,z)$, 
which is the same as 2D QCD on $t$-$z$ plane. 
In fact, by the gauge symmetry 
with $\Omega(t,z)$, 
gluon fields transform as
\begin{eqnarray}
    \label{eq:res_sym_tz}
    A_{t,z}(s) 
    & \to &
    \Omega(t,z)
    \left( 
    A_{t,z}(s) + \frac{1}{ig}\partial_{t,z}
    \right)
    \Omega^{\dagger}(t,z) , \\
    \label{eq:res_sym_xy}
    A_{x,y}(s) 
    & \to &
    \Omega(t,z)
    A_{x,y}(s)
    \Omega^{\dagger}(t,z), 
\end{eqnarray}
with keeping $R_{\mathrm{DR}}$ invariant.
From the gauge transformation \eqref{eq:res_sym_tz}, $A_{t}(s)$ and $A_{z}(s)$ behave like the gauge fields in 2D QCD, while $A_{x}(s)$ and $A_{y}(s)$ behave as charged matter fields which belong to the adjoint representation in 2D $\mathrm{SU}(N_{c})$ gauge theory.
Thus, DR-gauged QCD can be regarded as a 2D $\mathrm{SU}(N_{c})$ gauge theory with gauge fields $A_{t,z}(s)$ and charged matter fields $A_{x,y}(s)$.

The local condition of the DR gauge fixing 
is derived from the minimal condition of Eq.\eqref{eq:DRG} for gauge transformation, 
\begin{eqnarray}
    \label{eq:DRG_local}
    \partial_{x}A_{x}(s) + \partial_{y}A_{y}(s) = 0 ,
\end{eqnarray}
which is similar to that of the Landau gauge or the Coulomb gauge.
By introducing the DR-gauge fixing term  corresponding to Eq.~(\ref{eq:DRG_local}), 
the action of DR-gauged QCD is written as 
\begin{eqnarray}
    \label{eq:DRG_action}
    S_{\mathrm{DR}} &=& 
    \int d^{4} s~
    \Bigg[-
    \frac{1}{2}
    \mathrm{Tr} \; G_{\mu\nu}G^{\mu\nu}
    +
    \bar q(i\Slash{D}-m)q \cr
    &+&
    \frac{1}{2\alpha}
    \sum_{\perp = x, y}
    \mathrm{Tr}
    \left( \partial_{\perp}A_{\perp} \right)^{2}\Bigg],
\end{eqnarray}
using the DR-gauge fixing parameter $\alpha$.
The DR-gauged action $S_{\rm DR}$
has a residual gauge symmetry for $\Omega(t,z)$, which is the same as that 2D QCD on the $t$-$z$ plane.
In the path-integral formalism, 
the generating functional of DR-gauged QCD 
includes the Faddeev-Popov (FP) determinant, 
\begin{eqnarray}
\Delta_{\rm FP}\equiv {\rm Det}~(\partial_x D_x+\partial_y D_y),
\end{eqnarray}
which leads to the FP ghost interaction, 
\begin{eqnarray}
{\cal L}_{\rm ~FP} \equiv \bar c(\partial_x D_x+\partial_y D_y)c,    
\end{eqnarray}
in the local interaction form.
In fact, the DR-gauge fixing part 
includes only $A_x$ and $A_y$, 
and the FP (anti-)ghost fields 
directly couple only to them.
In contrast, the form of the action for the 
 $A_t$ and $A_z$ sector remains unchanged by the DR gauge fixing \cite{TS24}. 
 
\subsection{$tz$-projection and $xy$-projection}
\label{subsec:tz}

As shown in the previous study 
using lattice QCD \cite{TS24}, 
the amplitudes of 
$A_x$ and $A_y$ are strongly suppressed 
in the DR gauge. 
We then introduce the ``$tz$-projection'' 
or ``$A_x=A_y=0$ projection'', 
defined by setting the transverse gluon fields $A_{x}(s)$ and $A_{y}(s)$ to zero:
\begin{eqnarray}
    \label{eq:tz_proj}
    A_{x,y}(s) \to 0 \quad \hbox{($tz$-projection)}
\end{eqnarray}
with $A_{t,z}(s)$ remaining unchanged. 
By the $tz$-projection \eqref{eq:tz_proj}, 
the DR-gauged QCD action \eqref{eq:DRG_action} formally becomes 
the following tree-level expression:
\begin{eqnarray}
    \label{eq:DRG_act_tzp}
\!\!\!\!\!\! 
    S^{tz}_{\mathrm{DR}} \!\!\!\!\!\!
    &=& \!\!\!\!\!\!
    \int dxdy \!\!
    \int dtdz~
    \Bigg[ 
    \mathrm{Tr} \; G^{2}_{tz} 
    + 
    \sum_{\perp=x,y}
    \mathrm{Tr} 
    \left\{
      \left( \partial_{\perp}A_{t} \right)^2
    - \left( \partial_{\perp}A_{z} \right)^2
    \right\} \Bigg] \cr
    &+&
    \int d^4s~ \bar q\left(i\gamma_t D_t 
-i\gamma_z D_z
-i\gamma_x \partial_x
-i\gamma_y \partial_y
-m\right)q.
\end{eqnarray}
In the gauge sector, the DR-gauged YM theory after the $tz$-projection can be viewed as an ensemble of 2D YM-like systems on $t$-$z$ planes stacked along the $x$ and $y$ directions, interacting with each other \cite{TS24}.
In the quark sector, even after the $tz$-projection, the 4D structure is kept because of the presence of the 4D kinetic term of quarks.

Complementarily, 
we also introduce the ``$xy$-projection'' 
or ``$A_t=A_z=0$ projection'', 
defined by setting the transverse gluon fields $A_{t}(s)$ and $A_{z}(s)$ to zero:
\begin{eqnarray}
    \label{eq:tz_proj}
    A_{t,z}(s) \to 0 \quad \hbox{($xy$-projection)}
\end{eqnarray}
with $A_{x,y}(s)$ remaining unchanged.

\subsection{DR gauge and $tz$-projection in lattice QCD}
\label{subsec:DRG_lat}

In this subsection, 
we define the DR gauge and the $tz$-projection on a 4D lattice with spacing $a$ in the Euclidean space-time.
(For details, see Ref.~\cite{TS24}.)

In lattice QCD, the gauge variable is described with the link variable $U_{\mu}(s) \equiv e^{iag A_{\mu}(s)} \in \mathrm{SU}(N_{c})$, and 
the DR gauge is defined so as to maximize 
\begin{eqnarray}
    \label{eq:DRG_lat}
    R^{\mathrm{lat}}_{\mathrm{DR}}[U_x, U_y]
     \equiv 
    \sum_{s}
    \mathrm{Re Tr}
    \left[
        U_{x}(s) + U_{y}(s)
    \right]
\label{eq:RDRlat}
\end{eqnarray}
with gauge transformation.

In the lattice formalism, 
the $tz$-projection is defined as
\begin{equation}
    \label{eq:tz_proj_lat}
    U_{x,y}(s) \to 1 \quad \hbox{($tz$-projection)}
\end{equation}
with $U_{t,z}(s)$ remaining unchanged. 
Complementarily, 
the $xy$-projection is defined as
\begin{equation}
    \label{eq:xy_proj_lat}
    U_{t,z}(s) \to 1 \quad \hbox{($xy$-projection)}
\end{equation}
with $U_{x,y}(s)$ remaining unchanged. 

\section{Chiral symmetry breaking and Dirac modes}
\label{sec:chiral}

We briefly review 
spontaneous chiral-symmetry breaking and 
the Dirac modes in the SU($N_{\rm c}$) lattice QCD. 
In Euclidean QCD, we define 
all the $\gamma$-matrices to be hermite, 
i.e., $\gamma^\dagger_\mu=\gamma_\mu$ for 
$\mu=1,2,3,4,5$.

\subsection{Spontaneous chiral-symmetry breaking}

The 4D QCD action (\ref{eq:QCD_action}) 
has a global chiral symmetry 
${\rm SU}(N_f)_L \times {\rm SU}(N_f)_R$, 
reflecting the small current mass 
$m$ in the light quark sector with $N_f$ flavors.
The chiral symmetry is exact
in the chiral limit of $m=0$, 
although it is an approximate symmetry 
for $m \ne 0$, which introduces 
explicit breaking.

In the nonperturbative QCD vacuum, 
the chiral symmetry 
is spontaneously broken, 
and the chiral condensate 
$\langle\bar qq\rangle$ 
is a typical order parameter of 
chiral symmetry breaking.
Spontaneous chiral-symmetry breaking leads to the appearance of quasi-massless pions as Nambu-Goldstone bosons. 

\subsection{Dirac modes and the Banks-Casher relation}

Next, we consider the Dirac modes, i.e., 
the Dirac eigenstates of the Dirac operator 
$\Slash{D}$, since 
spontaneous chiral-symmetry breaking 
is directly related to 
the low-lying Dirac modes, 
as shown in the Banks-Casher relation \cite{BC80}. 
The Dirac eigenvalue equation 
is expressed as 
\begin{eqnarray}
\slashb{D}|n\rangle =i\lambda_n|n \rangle
\qquad (\lambda_n \in {\bf R})
\end{eqnarray}
with the Dirac eigenstate $|n \rangle$. 
Here, the Dirac eigenvalue $\lambda_n$ is real, since the Dirac operator $\Slash{D}$ is anti-hermite due to the hermite definition of $\gamma_\mu$. 
The set of normalized Dirac eigenstates is complete: 
\begin{eqnarray}
\sum_n|n \rangle\langle n|=1.
\end{eqnarray}
According to $\{\slashb{D},\gamma_5\}=0$,
the chiral partner $\gamma_5|n\rangle$ is also 
an eigenstate of $\Slash{D}$ with the eigenvalue $-i\lambda_n$.

In Euclidean QCD, 
the chiral condensate $\langle \bar{q}q \rangle$ is expressed as 
\begin{eqnarray}
 \langle \bar qq\rangle
=-\frac{1}{V}{\rm Tr}\frac{1}{\slashb D+m}
=-\frac{1}{V}\sum_n\frac{1}{i\lambda_n+m}, 
\label{qbarq}
\end{eqnarray}
with the space-time volume $V$ and 
the current quark mass $m$. 
From Eq.(\ref{qbarq}), 
the Banks-Casher relation \cite{BC80}
is derived,
\begin{align}
\langle\bar{q}q\rangle=-\lim_{m\to0}\lim_{V\to\infty}\pi\langle\rho(0)\rangle, \label{BCrel} 
\end{align}
where the Dirac eigenvalue density $\rho(\lambda)$ is defined by 
\begin{align}
\rho(\lambda)\equiv \frac{1}{V}\sum_n\langle\delta(\lambda-\lambda_n)\rangle.
\label{eq:density} 
\end{align}
From Eq.(\ref{BCrel}), the chiral condensate is proportional to the Dirac zero-eigenvalue density $\rho(0)$. 
Therefore, low-lying Dirac modes with small eigenvalues of $|\lambda_n|$ 
are essential 
for chiral symmetry breaking. 

In lattice QCD, 
the Dirac operator $\Slash{D}$ 
is expressed as a huge matrix: 
\begin{align}
 \slashb{D}_{s,s'} 
      = \frac{1}{2a} \sum_{\mu=1}^4 \gamma_\mu 
\left[ U_\mu(s) \delta_{s+\hat{\mu},s'}
        - U_{-\mu}(s) \delta_{s-\hat{\mu},s'} \right] \label{DiracOpExp}
\end{align}
with $U_{-\mu}(s)\equiv U^\dagger_\mu(s-\hat\mu)$.
In practical lattice calculations, 
we use the Kogut-Susskind (KS) formalism \cite{KS75,DSI14},  
which corresponds to the four-flavor case \cite{Rothe}.

\section{Numerical results}
\label{sec:numerical}

In this Letter, we perform 4D SU(3) lattice QCD Monte Carlo calculations at the quenched level. 
We use the standard plaquette gauge action 
with $\beta = 6.0$ and $16^4$, and 
we adopt the KS fermion
for the quark sector. 
The lattice spacing $a$ 
is determined such that the string tension $\sigma={\rm 0.89~GeV/fm}$
\cite{TS02}, and it is evaluated as 
$a \simeq 0.10~\mathrm{fm}$, i.e., $a^{-1} \simeq 2.0~\mathrm{GeV}$ at $\beta = 6.0$.
Using the pseudo-heat bath algorithm, we generate 50 gauge configurations, 
which are picked up every 1000 sweeps after 20,000 sweeps for thermalization.

\subsection{DR gauge fixing condition in lattice QCD}

After generating the gauge configurations, we perform DR gauge fixing \cite{TS24}.
For each gauge configuration, we numerically maximize  $R^{\mathrm{lat}}_{\mathrm{DR}}$ in Eq.\eqref{eq:DRG_lat}  
using an iterative maximization algorithm, 
similarly in Landau or maximally Abelian gauge fixing \cite{Iritani, Sakumichi}. 
For rapid convergence, we use the over-relaxation (OR) method 
with the OR parameter $1.6$. 
When $R^{\mathrm{lat}}_{\mathrm{DR}}$ is maximized, 
\begin{equation}
    \label{eq:gfix_violation}
    \Delta_{\mathrm{DR}}(s)
    \equiv 
    \sum_{\perp = x, y}
    \partial^{\mathrm{B}}_{\perp} \mathcal{A}_{\perp}(s)
\end{equation}
has to be zero.
Here, the lattice gluon field is defined as 
\begin{equation}
    \label{eq:lat_gauge_field}
    \mathcal{A}_\mu(s)
    \equiv 
    \frac{1}{2iag} 
    \left.
    \left[
        U_{\mu}(s) - U^{\dagger}_{\mu}(s) 
    \right]
    \right|_{\mathrm{traceless}} 
    \in
    \mathfrak{su}(N_{c}),
\end{equation}
where ``traceless'' means the subtraction of its trace part. 
As a numerical convergence criterion, we impose that the maximization of $R^{\mathrm{lat}}_{\mathrm{DR}}$ stops when 
$\epsilon_{\mathrm{DR}} < 4.0 \times 10^{-12}$ is satisfied
for 
\begin{equation}
    \label{eq:gfix_violation_norm}
    \epsilon_{\mathrm{DR}}
    \equiv 
    \frac{1}{N_{c} N_{s}}
    \sum_{s}
    \mathrm{Tr}
    \left[
        \Delta_{\mathrm{DR}}(s)
        \Delta^{\dagger}_{\mathrm{DR}}(s)   
    \right].
\end{equation}

\subsection{Dirac eigenvalues 
and low-lying Dirac-eigenvalue density}
\label{subsec:DiracEigen}
To extract the essential degrees of freedom for spontaneous chiral-symmetry breaking, 
we examine the Dirac-mode density $\rho(\lambda)$ in lattice QCD.
Using the KS fermion, 
we numerically calculate 
the low-lying eigenvalue $\lambda$ 
of the Dirac operator $\Slash{D}$ 
for each gauge configuration 
in the following three categories:
\begin{enumerate}
\item[(a)] 
original link-variables, 
\item[(b)] 
$tz$-projected link-variables in the DR gauge, 
\item[(c)]
$xy$-projected link-variables in the DR gauge.
\end{enumerate}
We use the shift-inverse Lanczos method 
for the calculation.

Figure~1 shows 
the total histogram of the 
low-lying Dirac eigenvalue $\lambda$ 
of the KS fermion for 50 gauge configurations for the three categories:
(a) original link-variables, 
(b) $tz$-projected link-variables, and 
(c) $xy$-projected link-variables.
For each figure, we calculate about 
500 low-lying Dirac eigenvalues 
with small $|\lambda_n|$ 
for each gauge configuration. 
For the visibility, 
the bin size is set 
to include about 100 data points 
in each bin of each figure, that is, 
$\Delta \lambda
=0.00128~a^{-1}$
is set in Figs.~1(a) and (b), 
and 
$\Delta \lambda= 0.0016~a^{-1}$
in Fig.~1(c).

\begin{figure}[h]
\centering
\includegraphics[width=8.0cm]{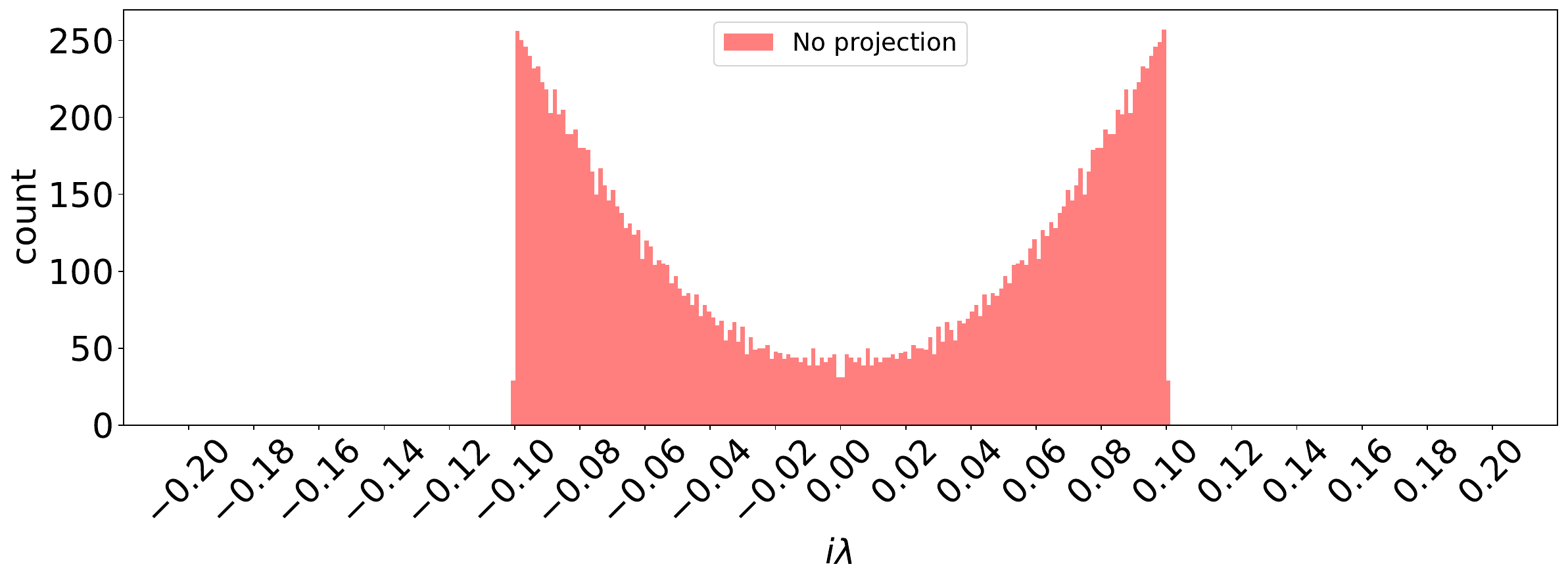}
\includegraphics[width=8.0cm]{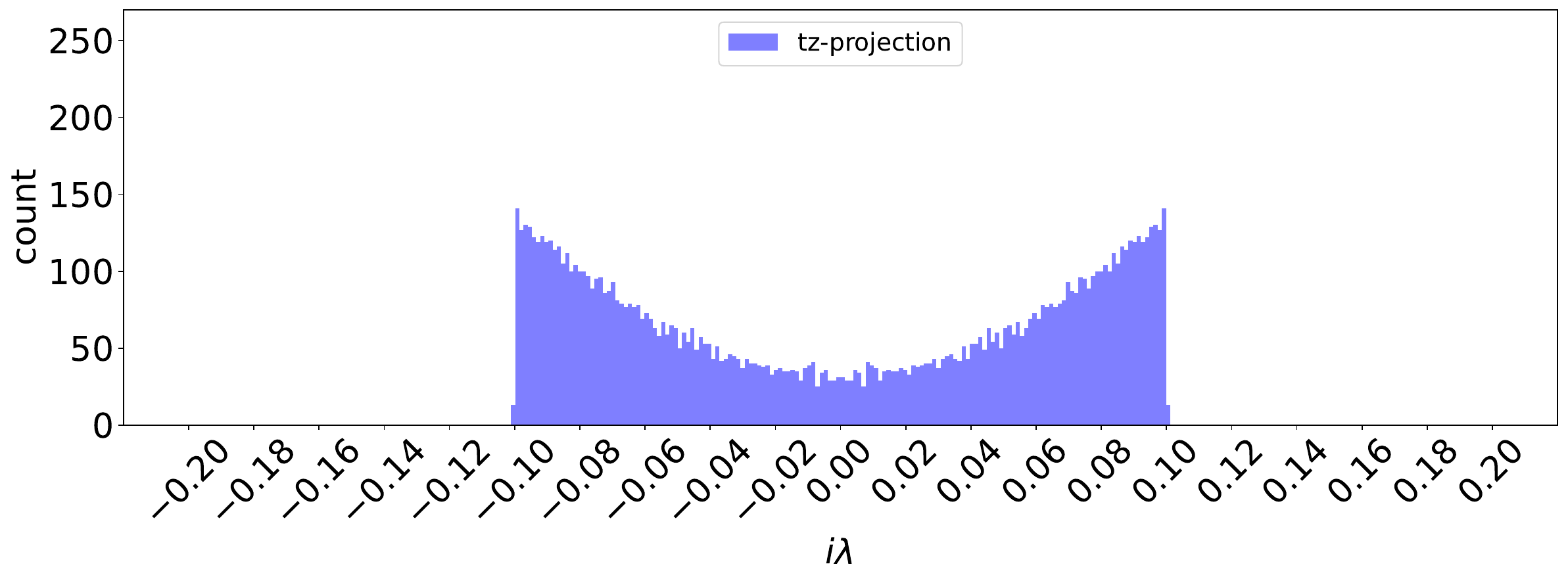}
\includegraphics[width=8.0cm]{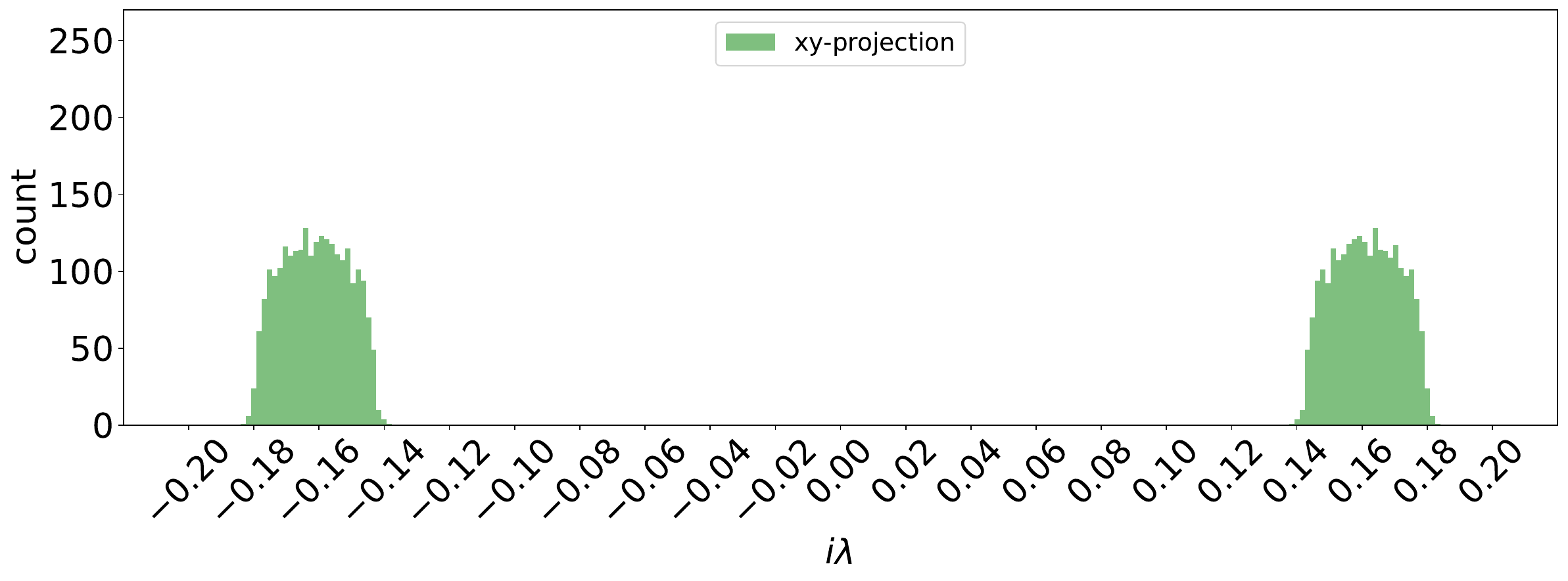}
\label{subcap:XYproj}
\caption{
The histograms of 
low-lying Dirac eigenvalues $\lambda$ for 
50 gauge configurations: 
(a) the original (non-projected) gauge configurations,
(b) the $tz$-projected DR gauge configurations, and
(c) the $xy$-projected DR gauge configurations.
They are proportional to 
the low-lying Dirac-eigenvalue densities,  
$\rho(\lambda)$, 
$\rho^{tz}(\lambda)$, 
and  
$\rho^{xy}(\lambda)$,  
respectively.
}
    \label{fig:full_tzp_hist}
\end{figure}

We show in Fig.~2 the low-lying Dirac-eigenvalue densities per flavor, 
$\rho(\lambda)$, 
$\rho^{tz}(\lambda)$ and $\rho^{xy}(\lambda)$,  
for the original, $tz$-projected, and $xy$-projected 
gauge configurations in the DR gauge, respectively.
For each of $\rho(\lambda)$ and $\rho^{tz}(\lambda)$, 
we add a parabola fitted to 
the low-lying region of  
$|\lambda_n| \le \lambda_{\rm IR\text{-}max}
\simeq 150~{\rm MeV}$, to  
estimate $\rho(0)$ and $\rho^{tz}(0)$  
at zero Dirac eigenvalue, $\lambda=0$.

\begin{figure}[h]
\begin{center}
    \includegraphics[width=8.0cm]{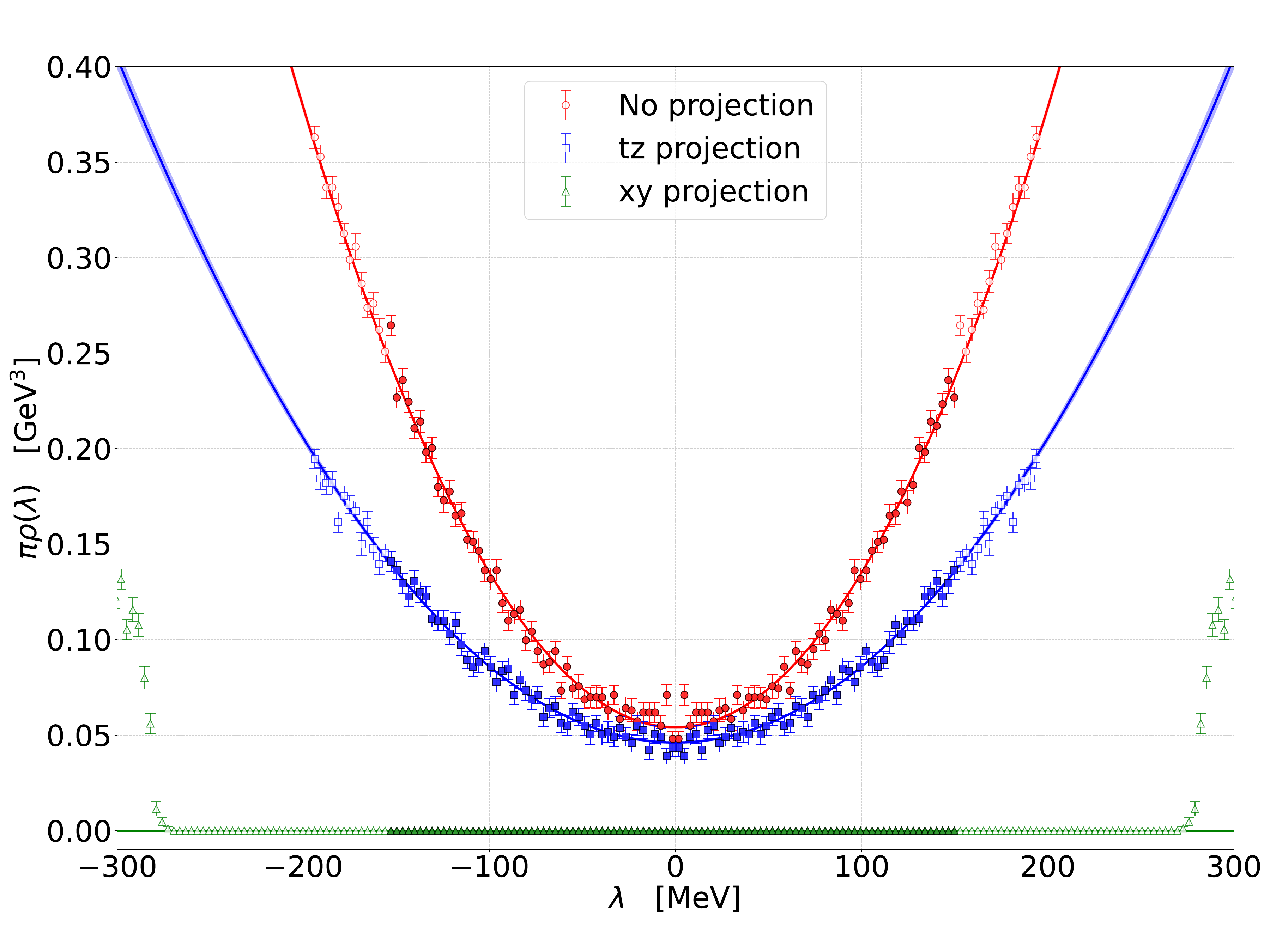}       
\end{center} 
\vspace{-15pt}
\caption{
The low-lying Dirac-eigenvalue densities 
per flavor, 
$\rho(\lambda)$, 
$\rho^{tz}(\lambda)$ and $\rho^{xy}(\lambda)$,  
for the original, $tz$-projected, and $xy$-projected 
gauge configurations in the DR gauge, respectively.
For each of $\rho(\lambda)$ and $\rho^{tz}(\lambda)$, 
a parabola fitted to the low-lying region is added.  
Filled markers represent data used for a parabola fit.
    }
    \label{fig:xyp_hist}
\end{figure}

\vspace{10pt}
\begin{figure}[h]
\begin{center}
    \includegraphics[width=8.0cm]{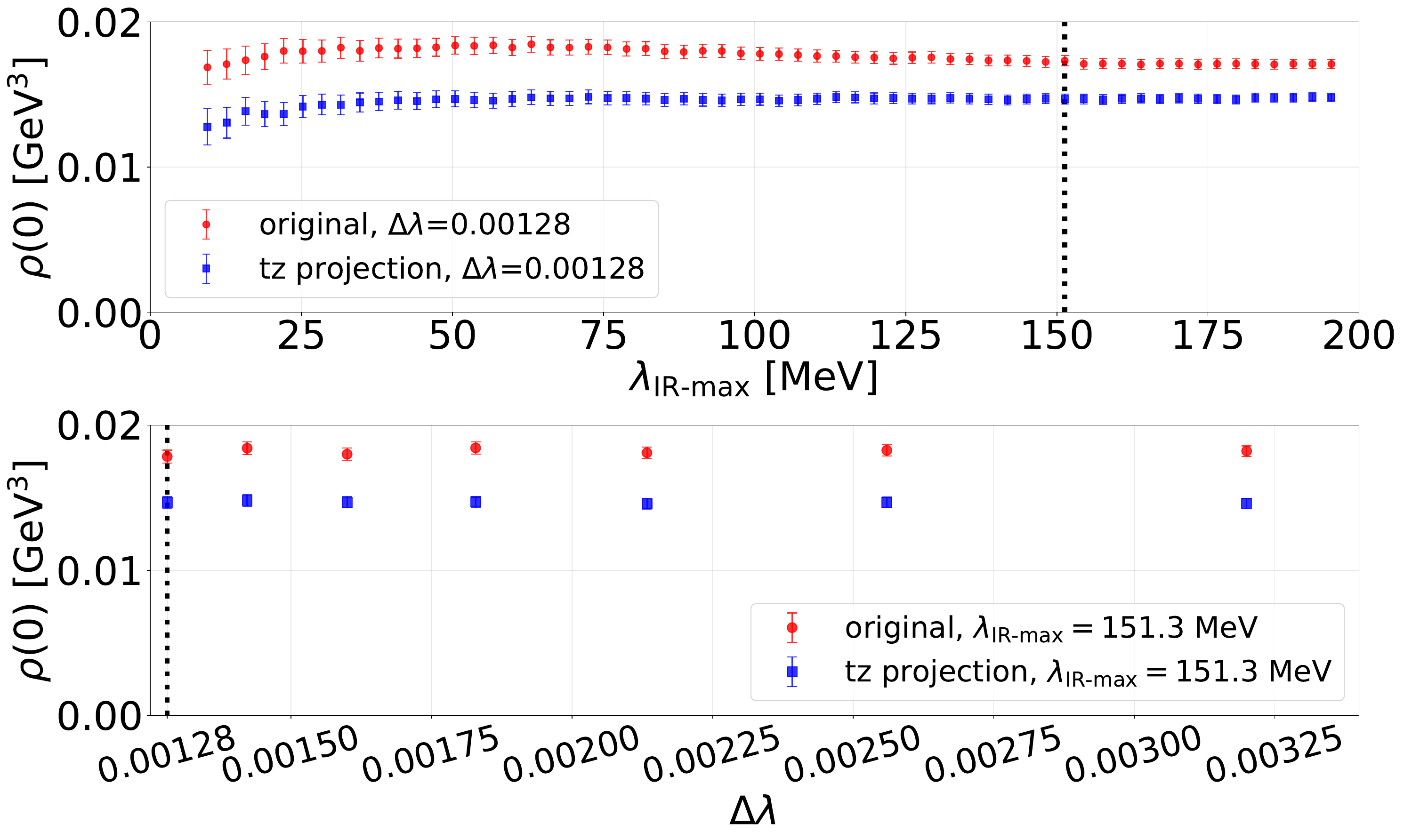}       
\end{center} 
\vspace{-15pt}
\caption{
The fit-range and bin-size dependence of the 
Dirac eigenvalue density $\rho(0)$ 
and the $tz$-projected value $\rho^{tz}(0)$.
The upper figure plots $\rho(0)$ 
against the maximum Dirac-eigenvalue 
$\lambda_{\rm IR\text{-}max}$ used for parabola fitting. 
The lower figure plots $\rho(0)$ 
against the bin size $\Delta \lambda$.
The dotted vertical lines represent 
the values used 
for the parabola fitting in
Fig.~\ref{fig:xyp_hist}.
}
    \label{fig:fit_bin_dep}
\end{figure}

From Fig.~2 and the parabola fit, 
we numerically estimate 
the Dirac-eigenvalue density $\rho(0)$, 
its $tz$-projected value $\rho_{tz}(0)$ and 
its $xy$-projected value $\rho_{xy}(0)$
at the zero eigenvalue $\lambda=0$:
\begin{eqnarray}
\rho(0) &\simeq& 0.00225 a^{-3} \simeq 0.01722~{\rm GeV}^{3}, \cr
\rho^{tz}(0) &\simeq& 0.00191 a^{-3} \simeq 0.01461~{\rm GeV}^{3}, \cr
\rho^{xy}(0)&=&0.
\end{eqnarray}
Since the QCD system is highly non-linear, it is not surprising that $\rho^{tz}(\lambda)+\rho^{xy}(\lambda)$ does not equal $\rho(\lambda)$.

The first remark is 
\begin{eqnarray}
\rho^{tz}(0)\simeq 0.85 \ \rho(0),
\end{eqnarray}
as a result of the $tz$-projection 
in the DR gauge.
This indicates that 
the $A_t$ and $A_z$ components carry 
the dominant contribution to 
spontaneous chiral-symmetry breaking
in the DR gauge.

The second remark is   
\begin{eqnarray}
\rho^{xy}(0)= 0
\end{eqnarray}
after the $xy$ projection in the DR gauge.
Thus, no spontaneous chiral-symmetry breaking is brought only with 
the $A_x$ and $A_y$ components  
in the DR gauge.

To ensure the numerical stability of 
the above estimates of 
$\rho(0)$ and $\rho^{tz}(0)$, 
we examine the dependence of the bin size $\Delta \lambda$  
and the fit region $\lambda_{\rm IR\text{-}max}$ 
of the parabola fit.
Figure ~3 (a) and (b) show the values of 
$\rho(0)$ and $\rho^{tz}(0)$ 
for various values of $\Delta \lambda$ and 
$\lambda_{\rm IR\text{-}max}$, respectively.
The dependence is found to be fairly small 
for both.

\subsection{Chiral Condensate}
\label{subsec:Chiral}

Now, we estimate 
the chiral condensate $\langle\bar qq \rangle$ per flavor, 
using the Banks-Casher relation (\ref{BCrel}), 
although 
the raw lattice estimate 
might be different from 
the standard value of 
the chiral condensate 
$\langle\bar qq \rangle \simeq -(250 {\rm MeV})^3$ 
at the renormalization point of $\mu=2{\rm GeV}$ 
in the $\overline{\rm MS}$ scheme.

Using the Banks-Casher relation (\ref{BCrel}), 
we estimate 
the chiral condensate $\langle\bar qq \rangle$, 
its $tz$-projected value $\langle\bar qq \rangle_{tz}$ and 
its $xy$-projected value $\langle\bar qq \rangle_{xy}$: 
\begin{eqnarray}
\langle\bar qq \rangle &\sim& -\pi \rho(0)\simeq -(378~{\rm MeV})^3, \cr
\langle\bar qq \rangle_{tz} &\sim& -\pi \rho^{tz}(0)\simeq -(358~{\rm MeV})^3, \cr
\langle\bar qq \rangle_{xy} &=&0.
\end{eqnarray}

Note again that the chiral condensate 
is largely preserved 
after the $tz$-projection
in the DR gauge, 
while it becomes zero 
after the $xy$-projection 
in the DR gauge.
In fact, 
only $A_t$ and $A_z$ are responsible to  
spontaneous chiral-symmetry breaking 
in DR-gauged QCD configurations. 

\section{Discussion}
\label{sec:discussion}

In this section, we discuss the implications of the aforementioned results on spontaneous chiral-symmetry breaking in the context of the physical picture of the QCD vacuum. 

\subsection{Quark confinement and spontaneous chiral symmetry breaking in DR-gauged QCD}

Through the analysis of gauge configurations generated in lattice QCD, we have demonstrated that only $A_t$ and $A_z$ are responsible for spontaneous chiral-symmetry breaking in DR-gauged QCD configurations. 

In the previous paper \cite{TS24}, 
only $A_t$ and $A_z$ are found to 
reproduce quark confinement 
in DR-gauged QCD configurations. 

In fact, in DR-gauged QCD, 
for both quark confinement 
and spontaneous chiral-symmetry breaking, 
only $A_t$ and $A_z$ are essential, 
while $A_x$ and $A_y$ do not contribute. 

\subsection{Propagation of gluons and quarks in DR-gauge QCD}

As shown in the previous paper \cite{TS24}, 
in DR-gauged QCD, 
gluon propagation in the $x$ and $y$ directions is highly suppressed, 
and gluons are bounded on $t$-$z$ planes. 
This seems to be an effective dimensional reduction. 

In contrast, quarks are not bounded on the $t$-$z$ plane, even in the DR gauge, 
because spontaneous chiral-symmetry breaking 
occurs in 4D QCD. 
Here, the Coleman theorem states that the realization of a real 2D system sharply contradicts the spontaneous breaking of global continuous symmetry and the resulting appearance of massless Nambu–Goldstone (pseudo-)scalar bosons.

In fact, the quark propagator 
in $tz$-projected Euclidean QCD takes the form,
\begin{eqnarray}
S^{tz}=\frac{1}{
i\gamma_t D_t+
i\gamma_z D_z+i\gamma_x\partial_x
+i\gamma_y\partial_y +m},
\end{eqnarray}
which clearly differs from the 2D propagator, 
\begin{eqnarray}
S^{\rm 2D}=\frac{1}{
i\gamma_t D_t+
i\gamma_z D_z+m},
\end{eqnarray}
reflecting the 4D quark kinetic term.
Spontaneous chiral-symmetry breaking 
is thus compatible with 
an effective two-dimensional description 
of quark confinement in the gluon sector 
in the DR gauge. 

In summary, 
as a result of spontaneous chiral-symmetry breaking in the chiral limit of $m=0$, 
massless pions appear 
as Nambu-Goldstone bosons. 
However, similar to quarks, 
the pions also propagate in 4D space-time.
This does not conflict with the Coleman theorem.

In this way, DR-gauged QCD system would be a system in which gluons (bosons) are bounded on $t$-$z$ planes and quarks (fermions) propagate between the planes.
This system is similar to the graphene \cite{graphene}, 
where electrons (fermions) are bounded on 2D planes and photons (bosons) propagate between planes.
Thus, the QCD system in the DR gauge might be regarded as the ``dual'' graphene, where roles of fermion and boson are interchanged.

\begin{figure}[h]
\begin{center}
    \includegraphics[width=7.0cm]{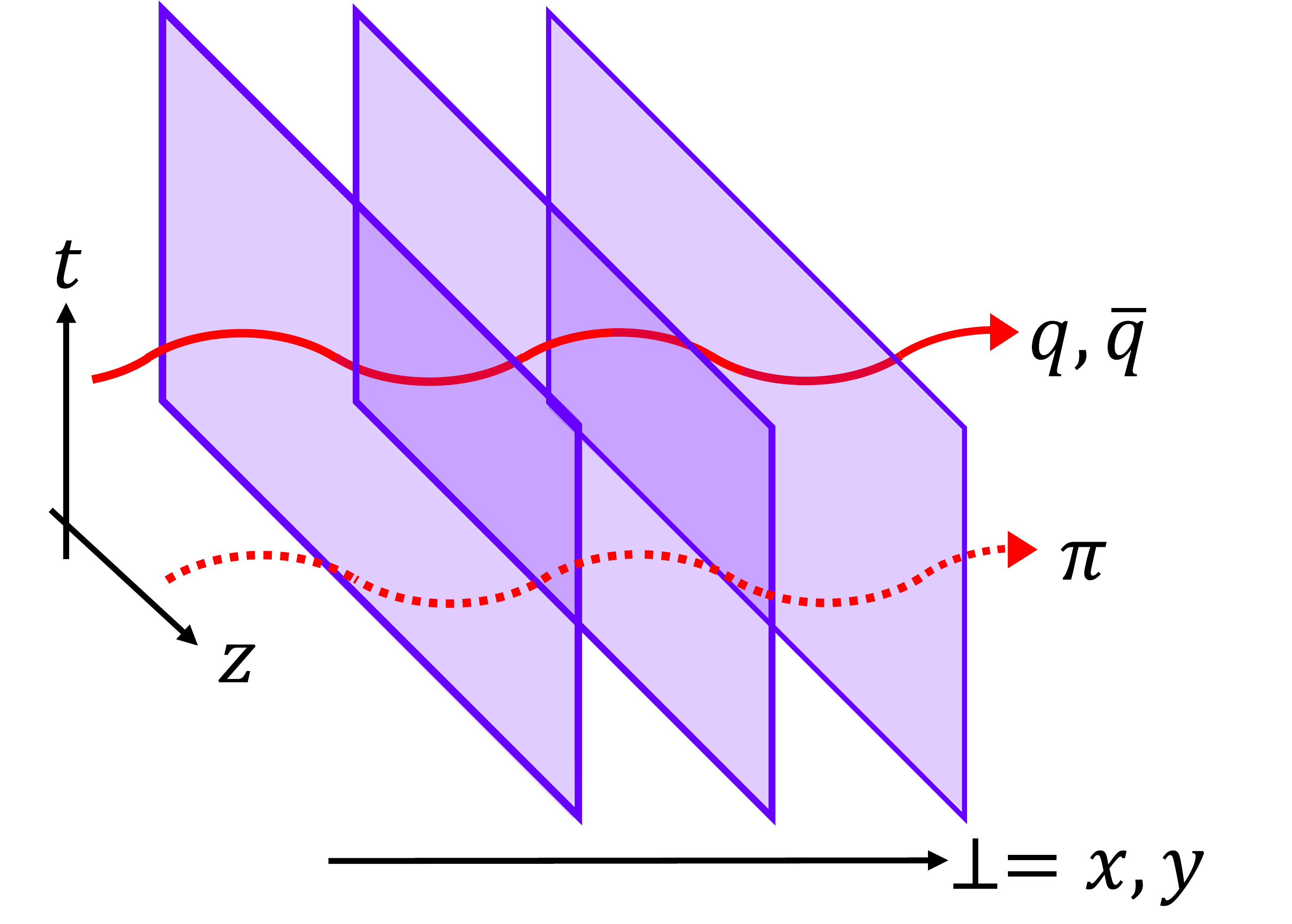}
\end{center}
\vspace{-10pt}
\caption{
    Schematic figure of the DR-gauged QCD 
    after the $tz$-projection.
    In the gluon sector, 2D QCD-like systems depicted as purple planes are piled in the $x$ and $y$ directions, and the gluon propagation is restricted within a short range in the $x$ and $y$ directions \cite{TS24}.
    In contrast, quarks and pions propagate over the 4D space-time. 
    }
    \label{fig:tz_proj.ed_QCD4_4d}
\end{figure}

\section{Summary and Concluding Remarks}
\label{sec:summary}

In this Letter, we have investigated spontaneous chiral-symmetry breaking in 4D lattice QCD 
in terms of the DR gauge and the $tz$- and $xy$-projections.
Here, the $tz$- and $xy$-projections 
have been performed for the gauge configurations 
generated in lattice QCD. 

The previous study \cite{TS24} indicates 
a dominant role for $A_{t}(s)$ and $A_{z}(s)$ 
in quark confinement in DR-gauged QCD,  
together with an effective dimensional reduction of the 4D Yang-Mills theory into an ensemble of 2D Yang-Mills-like systems on the $t$-$z$ plane.
In this work, we have examined whether the same dominant role of $A_{t}(s)$ and $A_{z}(s)$ also holds for spontaneous chiral-symmetry breaking.

To this end, 
we have calculated the low-lying eigenvalues 
of the Dirac operator $\Slash{D}$ 
for the $tz$- and $xy$-projected 
gauge configurations in the DR gauge 
in $\mathrm{SU}(3)$ quenched lattice QCD,  
using the Kogut-Susskind formalism. 

In the DR gauge, the distribution of low-lying Dirac eigenvalues is fairly well-preserved by the $tz$-projection, but lost by the $xy$-projection.
According to the Banks-Casher relation, 
the chiral condensate, the order parameter of spontaneous chiral-symmetry breaking, is approximately preserved by removing $A_{x}$ and $A_{y}$ in the DR gauge, but is completely lost by removing $A_{t}$ and $A_{z}$.

We have also argued that 
spontaneous chiral-symmetry breaking 
is compatible with the gluonic dimensional reduction in DR-gauged QCD 
because quarks and pions propagate 
in 4D space-time.

In conclusion, 
similar to quark confinement, $A_{t}$ and $A_{z}$ play a dominant role in spontaneous chiral-symmetry breaking in the DR gauge, while $A_{x}$ and $A_{y}$ do not.
This result suggests that the essential features of both quark confinement and spontaneous chiral-symmetry breaking in 4D QCD can be described only with $A_{t}$ and $A_{z}$ in the DR gauge.

As a future work, 
it is interesting to perform the lattice QCD calculations for various $\beta$ and lattice volumes, to investigate the scaling property in the DR gauge, and to extend the present analysis to full QCD with dynamical quarks and to finite temperature. 

\section*{Acknowledgement}
\label{sec:acknowledgement}
K.T. is supported by JSPS KAKENHI Grant Number 25KJ1618. 
The lattice QCD calculation was performed by SQUID at RCNP at Osaka University.
We use LAPACK and ARPACK 
to perform the Lanczos method.

\end{document}